# Coherent photon beam based diagnostics for a seeded extreme ultraviolet free-electron laser

Chao Feng, Haixiao Deng[*], Manzhou Zhang, Shunqiang Tian, Meng Zhang, Bo Liu, and Zhimin Dai

*Shanghai Institute of Applied Physics, Chinese Academy of Sciences, Shanghai, 201800, China*

**Abstract:**

Independently from electron beam based procedures, photon beam based diagnostics is an alternative way for alignment and commissioning of the numerous undulator cells in a high-gain short-wavelength free-electron laser (FEL). In this paper, using the seed laser modulated electron beam and the undulator fine tuning technique, a coherent photon beam based diagnostic was proposed for seeded FEL, and some preliminary experimental results at Shanghai deep ultraviolet FEL test facility were presented. It demonstrates that the spatial distribution of the coherent harmonic radiation from one individual or two consecutive undulator segments can be used to optimize the electron beam trajectory, to verify the undulator magnetic gap, and to adjust the phase match between two undulator segments.

* Corresponding author: denghaixiao@sinap.ac.cn

## 1. Introduction

With the great success of the world's first x-ray free-electron laser (FEL) [1], scientists now have the ability to investigate the matter with femtosecond x-ray single-shot diffraction [2] and time-resolved pump-probe experiments [3], etc. Therefore, more and more FEL user facilities are in operation and under construction all over the world, from the extreme ultraviolet [4-7] to hard x-ray [8-11] spectral region. Usually, short-wavelength FEL undulator system is made up of periodic undulator modules, which consists of an undulator segment and a long intersection containing various items such as cavity BPM, quadrupole, phase shifter, steering coils and vacuum components.

As well known, the FEL exponential gain requires stringent electron beam trajectory control along the whole undulator system, fine undulator magnetic field setting and well-matched phase between two consecutive undulator segments. For this purpose, the electron beam based alignment [12-14] and the photon beam based diagnostics [15-16] have been developed, respectively. While the former has already got great success in x-ray FEL [1], it shows no insight into the magnetic gap or the phase match of adjacent undulator segments. Photon beam based diagnostics serve as an additional tool for the beam alignment and FEL undulator commissioning, independently from the electron beam based procedure.

Currently, photon beam based diagnostics is routinely used for SACLA Facility [16] and seriously considered by European XFEL project [15], and had been proposed for the LCLS project [17]. The photon beam instruments consisting of a crystal monochromator, a spatial imaging optics and intensity detectors are located downstream of the undulator system, and thus allow to characterize the radiation of selected undulator segment and can be used for photon beam based alignment. The one common diagnostics facilitates a precise alignment and setup of the whole undulator system.

However, photon beam based diagnostics has only been considered for a self-amplified spontaneous emission (SASE) [18] x-ray FEL. In recent years, several seeded FEL user facilities are proposed from extreme ultraviolet to soft x-ray region [5-7] where the monochromic light is directly generated with the help of various laser-beam interactions [19-21]. In this paper, using the radiation from spatial density modulated electron beam and the undulator fine tuning technique, a modified, coherent photon beam based diagnostic namely, was proposed for seeded FEL. We first illustrated the principle to optimize the electron beam trajectory, to verify the magnetic gap, and to adjust the phase match between two successive undulator segments, on the basis of Dalian coherent light source [7]. Finally, preliminary coherent photon beam based diagnostics experimental results at Shanghai deep ultraviolet FEL test facility (SDUV-FEL) [22] were presented.

**2. Coherent photon beam based diagnostics**

The schematic of the seeded FEL system layout is shown in Figure 1. The electron beam is spatially modulated after the modulator undulator and the dispersive chicane. It indicates that the electron beam with strong bunching factor emits coherently at the radiator undulator. The intensity analysis of such coherent photon beam will illustrate the temporal structure of the electron beam current, emittance and energy spread [23-25], and the spatial profile study will allow optimizing the electron beam trajectory, verifying the undulator magnetic gap, and adjusting the phase match between two undulator segments.

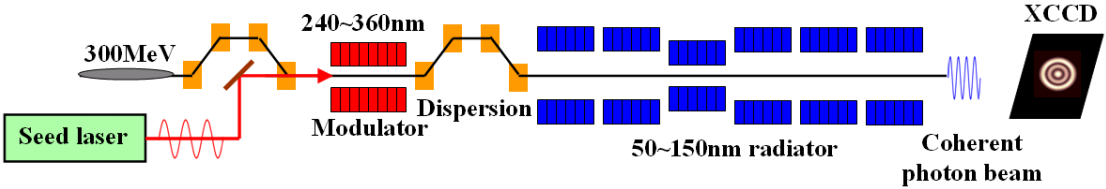

Figure 1. Schematic of a seeded extreme ultraviolet FEL user facility.

Generally, the basic idea of photon beam based diagnostics is the spatial distribution analysis of the spontaneous emission from radiator undulator, where a crystal monochromator is utilized for accuracy spectrum selection with respect to the undulator resonant wavelength. However, because of the laser induced bunching, only part of the undulator radiation spectrum is significantly amplified in a seeded FEL, usually the harmonics of the seed laser with relatively narrow bandwidth. As shown in Figure 2, since the electron beam strongly bunched at 50nm, the coherent undulator radiation results a monochromatic spectrum at 50nm, while the undulator resonant wavelength is changed in a reasonable range. Thus, with the coherent amplified photon spectrum fixed, the radiator spontaneous emission spectrum can be relatively shifted by varying the undulator gap. And then coherent photon beam based diagnostics can be accomplished without crystal monochromator.

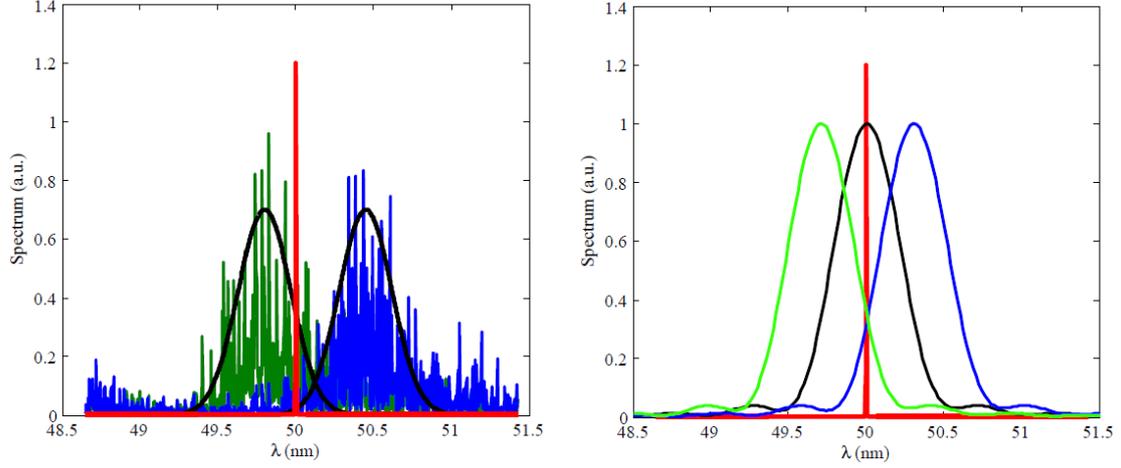

Figure 2. The red line represents the coherent photon beam from undulator, while others represent the undulator spontaneous emission. The spontaneous emission spectrum in the left plot is from GENESIS [26] simulation by turn off the seed laser, where the green means the undulator resonant wavelength $\lambda_R$ = 49.7nm, the blue means $\lambda_R$ = 50.5nm. The spontaneous emission spectrum in the right plot is from SPECTRA [27] calculation, where the green means $\lambda_R$ = 49.7nm, the black means $\lambda_R$ = 50.0nm and the blue means $\lambda_R$ = 50.3nm.

According to the undulator radiation physics, the spectral power per unit solid angle due to a single electron in an undulator leads to the following expressions:

$$\frac{\partial^2 P}{\partial \lambda \partial \Omega} \propto \left( \frac{\sin(\frac{\lambda - \lambda_R}{\lambda_R} \pi N)}{\frac{\lambda - \lambda_R}{\lambda_R} \pi N} \right)^2, \quad (1)$$

$$\lambda_R = \frac{\lambda_u (1 + K^2/2 + \gamma^2 \theta^2)}{2\gamma^2}, \quad (2)$$

$$K \approx 0.934 \, (1 + \frac{2\pi^2 y^2}{\lambda_u^2}) B_0[T] \lambda_u[cm], \quad (3)$$

where $\lambda_R$ is the fundamental resonant wavelength of the undulator radiation, $N$ is the undulator period number, $\lambda_u$ is the undulator period length, $K$ is the normalized undulator parameter, $\gamma$ is the kinetic energy of the electrons measured in units of its rest mass, $\theta$ is the observation angle, and $B_0$ is the peak magnetic field on the undulator axis.

While the electron beam is spatially bunched at 50nm, Figure 3 displays the normalized transverse distributions of coherent photon beam for slightly tuning of the fundamental undulator peak, where all calculations have been performed for an observation point 3m behind the undulator. The coherent radiation power scan result suggests an optimized radiator undulator resonant wavelength of 49.84nm. This small discrepancy is due to well known undulator detuning effect. It is clear to see that a Gaussian spatial mode is observed when the coherent photon beam power is at maximum. For undulator resonant wavelength shorter than 49.84nm, a broadening accompanied by splitting of the intensity cone towards rings is obtained. Detuning the undulator resonant wavelength by 0.36nm longer leads to a considerable narrowing of the radiation cone, however to the expense of lower intensity. These spatial distributions of coherent photon beam can be intensively analyzed with Eqs. (1) to (3), and thus used for electron trajectory alignment and commissioning of the FEL undulator modules.

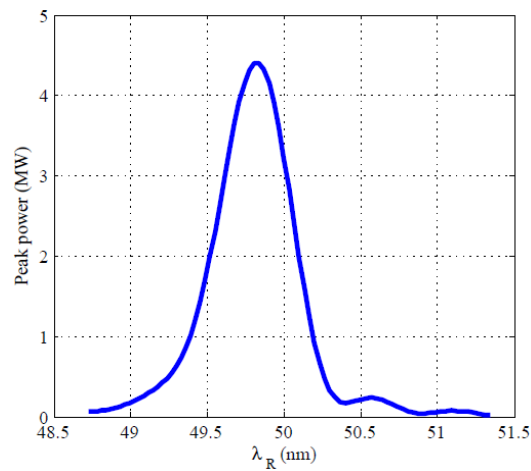

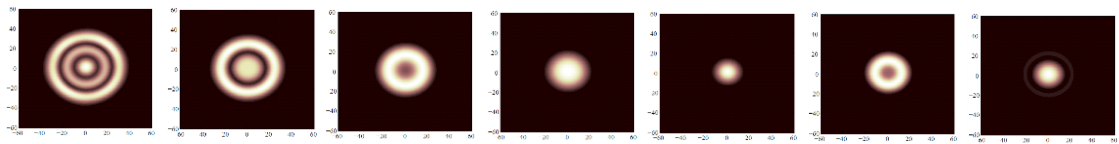

Figure 3. The peak power (the upper plot) and the spatial profile (the lower plots) of coherent photon beam dependence on the radiator resonant wavelength. The radiator resonant wavelength is 48.84nm, 49.14nm, 49.49nm, 49.84nm, 50.20nm, 50.47nm and 50.70nm from left to right respectively.

3. **Numerical calculations for Dalian coherent light source**

In this section, numerical examples of coherent photon beam based diagnostics is illustrated on the basis of recently proposed Dalian coherent light source [7], which is a seeded extreme ultraviolet FEL user facility. The linac of Dalian coherent light source will provide electron beam with the energy up to 300MeV, the bunch charge of 500pC, the pulse length of 1ps, and the normalized slice emittance lower than 1mm-mrad. With combination of optical parametrical amplification laser technique (240-360nm seed laser), variable gap undulator (30mm period length, 9-18mm alterable gap radiator) and harmonic selection (the $2^{nd}$ harmonic to the $5^{th}$ harmonic), Dalian coherent light source is able to generate fully coherent FEL radiation with continuous wavelength tuning ability over 50-150nm based on high-gain harmonic-generation (HGHG) [19] scheme. In the following calculation, only the 50nm output case is considered for simplification.

Figure 4 shows the start-to-end simulation results of Dalian coherent light source. It is clearly that the 50nm FEL radiation performs good temporal coherence and powerful output pulse energy. It is worth to stress that the undulator tapering technique [28-31] will be utilized to the last 2 undulator segments for enhancing the final output pulse energy by a factor about two. It means that the magnetic gap of 6 undulator segments should be independently tuned to the optimal.

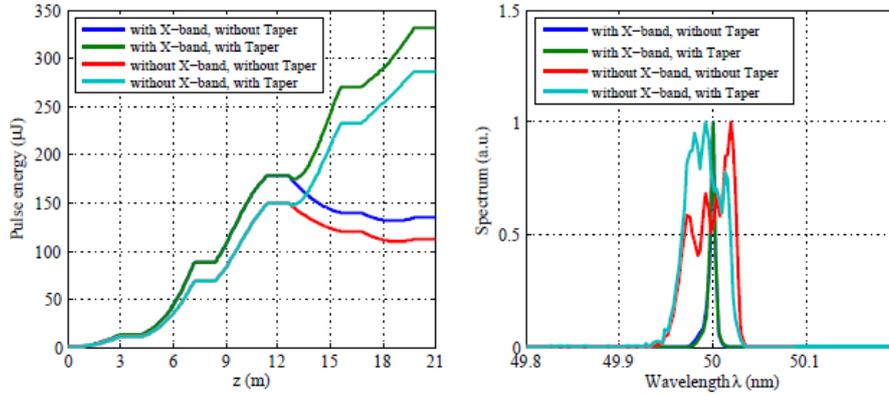

Figure 4. Start-to-end simulation results on FEL peak power growths in the radiator undulator and the final output radiation spectra of Dalian coherent light source.

Coherent photon beam based diagnostic provides an opportunity to quickly optimize the magnetic gap of the radiator undulator. If we suppose an observation point 3m after the final radiator undulator, gap tuning of 6 undulator segments can be accomplished in a successive way, i.e. always only one undulator gap is closed. Then the spatial distribution of coherent radiation at the observation point depends on the radiation undulator gap setting, as shown in Figure 5. Coherent photon beam of the first 4 radiator undulator segments shows an ideal Gaussian spatial distribution, since the radiator undulator resonance of them is optimized for maximum output intensity. Meanwhile, because a stepped taper is applied to the last 2 radiator segments, i.e., the undulator resonant wavelength become shorter, a ring shape rather than core shape spatial distribution of coherent photon beam will be observed.

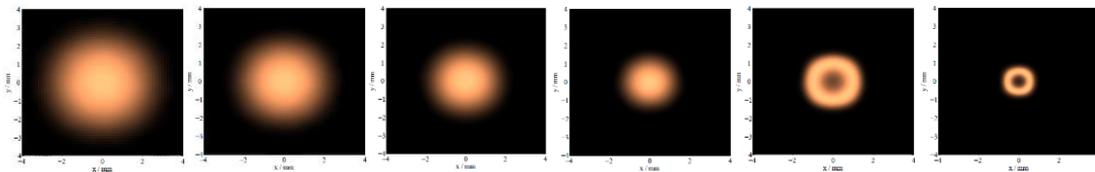

Figure 5. Coherent photon beam spatial profiles from 6 single radiator segments, where a stepped taper is applied to the last 2 radiator undulator for radiation power enhancement.

Generally, to compensate the wavelength dependent phase condition when changing the undulator magnetic gap, a phase shifter is installed between two adjacent radiator segments. It consists of a two periods' undulator with gap-dependent magnetic field and delays the electron beam so that the radiation from the following segment is in phase with that of the previous for all wavelengths. The adjustment of the phase has to be assured with an accuracy of a few degrees for Dalian coherent light source. The determination of the correct phase is based on observing the spatial distribution of the composed coherent photon beam of two successive radiator undulator segments (Figure 6). Here we consider two cases with undulator resonant wavelength of 49.49nm and 49.84nm, respectively. For a complete phase match, the spatial radiation distribution of the composed radiation is similar with the matched phase condition (upper), whereas the coherent photon beam is emitted in a ring in the fully destructively interfering case (lower).

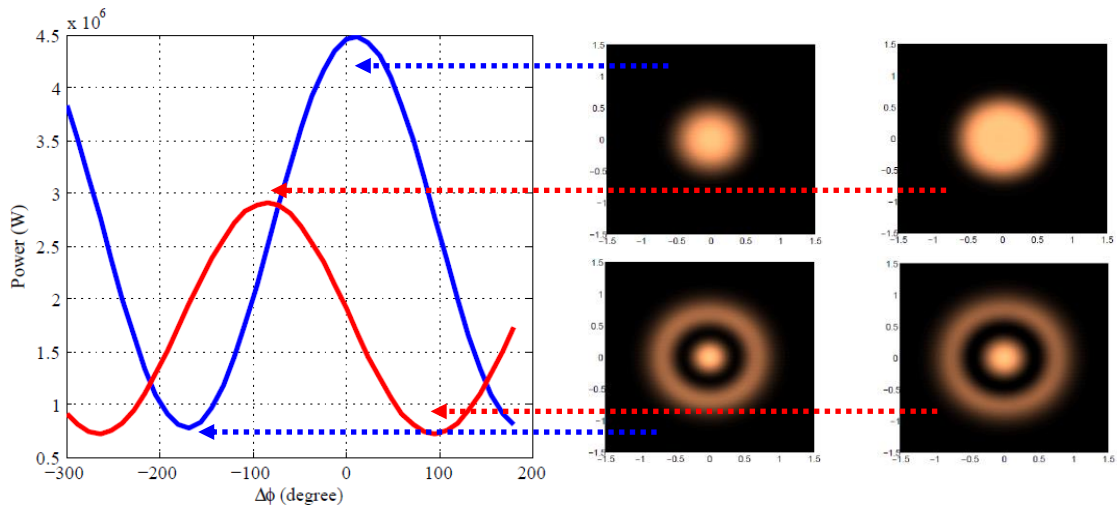

Figure 6. Coherent photon beam spatial distributions of two successive undulator segments (red: tuned to a wavelength of 49.49nm, blue: tuned to a wavelength of 49.84nm, spatial distribution from one undulator segment can be seen in Figure 3) for the best phase difference (the right upper two plots) and the worst phase difference (the right lower two plots).

It is well known that random quadrupole offset will lead to misalignment of electron beam trajectory along the undulator system, which leads to mismatch of FEL resonance and transverse overlap between the electron beam and FEL radiation, and thus results in the FEL gain reduction. For Dalian coherent light source, FEL physics study shows an upper limit trajectory displacement of 20μm and 10μrad angular trajectory alignment requirement within a single undulator segment. While the electron beam trajectory can be perfectly corrected by the electron beam based alignment, coherent photon beam based alignment will be an additional tool.

Figure 7 gives the spatial profiles of coherent photon beams when the electron beam enters a single radiator segment with a transverse offset or angle, the radiator resonant wavelength is set to be less than the coherent photon beam wavelength, i.e., 49.49nm in our case. The beam trajectory information can be obtained from the projection of the central part of spatial distribution. As shown by the first column plots in Figure 4, because of an initial electron beam offset in x-plane, the center of the coherent photon beam ring derivates from the undulator axis in horizontal, however, the projection amplitude does not change due to the relatively wide good field region of undulator in the x-plane. The phenomenon due to the beam offset in y-plane could be explained by Eq. (3) induced undulator parameters *K* enhancement. A tilted electron beam trajectory mainly changes the definition of the observation angle $\theta$ in Eq. (2), and thus introduces the variation shown in the last two columns.

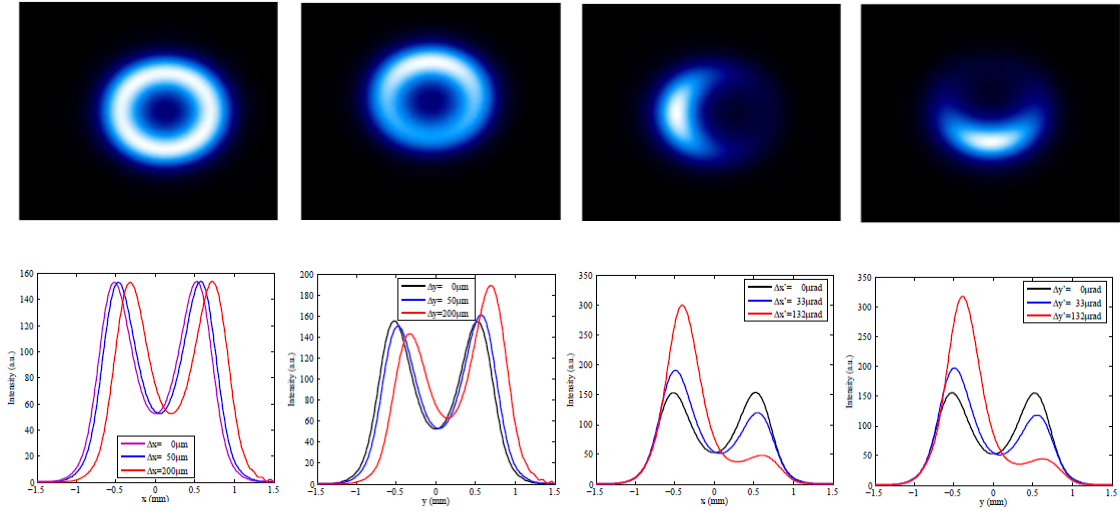

Figure 7. The spatial distributions (the upper) and the central part projection (the lower) of the coherent photon beam with controlled electron beam offset and tilted injection with respect to undulator axis.

## 4. Preliminary experiment at SDUV-FEL

The SDUV-FEL is a multi-purpose test facility for seeded FEL studies, capable of testing various FEL working modes, such as Cascaded HGHG [32-34], echo-enabled harmonic generation [35-38] and the crossed-planar undulator technique [39-40], etc. During the experiment of coherent photon beam diagnostics, the linac beam energy of the SDUV-FEL was set to be 148MeV. The seed laser comes from the commercial Ti-Sa system, which can provide up to several tens micro Joule energy with about 100fs pulse duration and 800nm wavelength. The modulator, a 10×50mm period length permanent magnet undulator, was set to fulfill the resonant conditions with beam energy and the seed laser. The radiator consisting of two segments of 1.6m long permanent magnet undulators with 40mm period length, and the magnetic gap of the radiator was tuned to be 16.5mm for resonant at the 2$^{nd}$ harmonic of the seed laser, i.e., 400nm. A visible CCD camera located downstream of the radiator was utilized for monitoring the spatial distribution of the coherent 400nm photon beam.

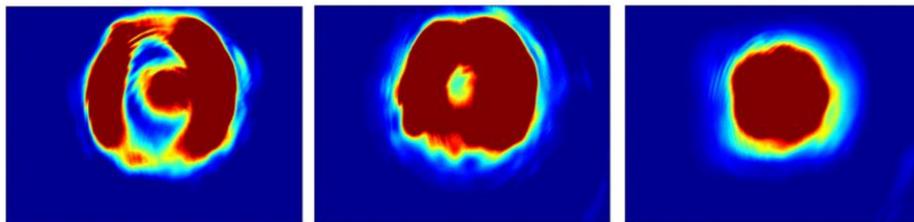

Figure 8. Spatial distributions of the 400nm coherent photon beam observed on the CCD camera downstream of the radiator.

Figure 8 illustrates the spatial distributions of the 400nm coherent radiation pulses when adjusting the magnetic gap of the radiator, where the ring shape and core shape coherent photon beam profiles were both observed on the CCD camera. The experimental phenomenon demonstrated the theoretical predictions as shown in Figure 3. And this phenomenon is successfully used to optimize the radiator gap to fit the resonant condition. It was found that the composed radiation intensity was on maximum when the spatial distribution of coherent photon beam from two single radiators transferred from a ring to a core profile, and transversely overlapped each other.

Figure 9 (left) recorded the spatial distribution of coherent photon beam when changing the entering angle of the electron beam at the entrance of the radiator. The radiator resonant wavelength was set to be a little bit less than the 400nm. Comparing with the spatial distribution in Figure 8 (middle), Figure 9 (left) indicates a misalignment of electron beam trajectory in the radiator. The start-to-end simulation gives a similar spatial distribution as shown in Figure 9 (right), where a kicker is used to horizontally kick the electron beam at the entrance of the radiator, and thus results in an electron beam central offset in x-plane at the exit of the radiator of about 400 μm.

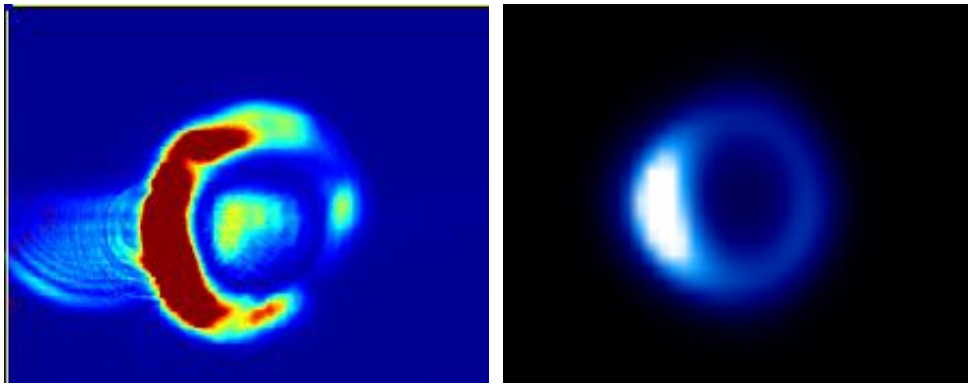

Figure 9. Spatial distribution of the 400 nm coherent photon beams observed on the CCD camera downstream of the radiator when the electron beam enters the radiator undulator with a tilt angle (left) and the corresponding start-to-end simulation result (right).

## 5. Conclusions

With the fully coherent properties and the intrinsic amplification advantages of the output radiation from a single radiator segment in a seeded FEL, in the absence of a crystal based monochromator, a coherent photon beam based diagnostics is proposed to optimize the electron beam trajectory, to verify the undulator magnetic gap, and to adjust the phase match between two undulator segments. The principle of the proposed coherent beam based diagnostics is illustrated with the parameters of Dalian coherent light source, the first FEL user facility in China. Moreover, preliminary experimental results at Shanghai deep ultraviolet FEL test facility are presented. It is worth to stress that, all the diagnostics in this paper is based on the spatial distribution analysis of coherent photon beam. If combined with the intensity investigation, coherent photon beam based diagnostics is an effective and a complementary method for commissioning and alignment of FEL undulator system, independently from the electron beam based procedures.

## Acknowledgements

The authors would like to thank T. Zhang, K. R. Ye, D. Wang and Z. T. Zhao for useful discussions. This work is partially supported by the Major State Basic Research Development Program of China (2011CB808300) and the National Natural Science Foundation of China (11175240, 11205234 and 11322550).